\documentclass[a4paper,preprintnumbers,amsmath,amssymb,nofootinbib,showpacs,10pt]{revtex4}
\usepackage[margin=2cm]{geometry}
\usepackage{amssymb}
\usepackage{mathrsfs}
\usepackage{graphics}
\def\P{{\cal{P}}}

\def\be{\begin{equation}}
\def\ee{\end{equation}}
\def\bea{\begin{eqnarray}}
\def\eea{\end{eqnarray}}

\newcommand\lsim{\mathrel{\rlap{\lower4pt\hbox{\hskip1pt$\sim$}}
    \raise1pt\hbox{$<$}}}
\newcommand\gsim{\mathrel{\rlap{\lower4pt\hbox{\hskip1pt$\sim$}}
    \raise1pt\hbox{$>$}}}

\begin{document}

\title{Possibly Large Corrections to the Inflationary Observables}
\author{N. Bartolo$^{1,2}$ and 
A. Riotto$^{2,3}$}

\affiliation{$^1$ Dipartimento di Fisica, Universit\'a di  Padova, Via Marzolo,
8 - I-35131 Padua -- Italy\\}

\affiliation{$^2$ INFN, Sezione di Padova, Via Marzolo,
8 - I-35131 Padua -- Italy\\}

\affiliation{$^3$ CERN, Theory Division, Gen\'eve 23, CH-1211 Switzerland\\}
% \date{\today}

\pacs{98.80.Cq  \hfill CERN-PH-TH/2007-230}

\begin{abstract}
\noindent
We point out  that the theoretical predictions for the inflationary observables may be generically altered 
by the presence of fields which are heavier than the Hubble rate during inflation and whose
dynamics is usually neglected. They introduce
corrections which may be easily larger than both the second-order contributions in the slow-roll
parameters and the accuracy expected in the
forthcoming experiments.

\end{abstract}

\maketitle
\noindent
Cosmological inflation \cite{lrreview} has become the dominant paradigm
within which one can attempt to
understand the initial conditions for Cosmic Microwave Background (CMB)
anisotropies and structure formation. In the inflationary picture, 
the primordial cosmological perturbations are created from quantum 
fluctuations which are
``redshifted'' out of the horizon during an early period of 
accelerated expansion.
Once outside the horizon, they remain ``frozen'' until the horizon
grows during a later radiation- or matter-dominated era.
After falling back inside the horizon they are communicated to the primordial
plasma and hence are directly
observable as temperature anisotropies in the CMB.
These anisotropies have been mapped 
with spectacular accuracy by the  Wilkinson Microwave 
Anisotropy Probe (WMAP)~\cite{wmap} and even a better accuracy will be reached
by  the Planck satellite \cite{planck} and its
successors. This has allowed to put more and more stringent constraints on inflationary 
models~\cite{wmap,KK1,martin,finelli}. 

Given the present and future level of accuracy of the cosmological observations in the
CMB anisotropies and in the large scale structure of the universe, the theoretical predictions for the
inflationary observables need to be as precise as possible. Within single field models
of inflation this is not necessarily a hard task since the power spectrum $\P_\zeta$ of the
curvature perturbation $\zeta$, the  spectral index $n_\zeta$, the amount of tensor modes and
the way these observables run with the scale may be computed with the required  level of accuracy
in terms of series of powers of the slow-roll parameters~\cite{Lidseyetal}. Comparing cosmological
data with the predictions of a given model of inflation to decide whether they are compatible
seems therefore quite straightforward. In this short note, we would like to point out
that this may not be the case: there are generic corrections to the theoretically  predicted inflationary
observables which have been neglected so far and which may, in principle, be calculated
within a given inflationary model once the latter is rooted in a well-defined particle
physics model. 

Let us briefly explain the nature of such corrections. In any inflationary
model it is a common lore to neglect the dynamics of those fields which are heavier than
the Hubble rate $H$ during inflation. This is because the heavy fields are  stabilized
at the minimum of their potential and are thought to  play no active role in the inflationary dynamics. 
This is certainly correct, they do not play any major role. However, the correct question
is if they influence the inflationary predictions at the level of accuracy needed for a fair
comparison with the observations. We believe the answer may be positive. 
This may be either fortunate or unfortunate according to the various cases. The key point
is that during inflation the vacuum expectation value (VEV) of these heavy fields is not likely
to be constant; on the contrary it is expected to adiabatically and slowly changing with time
to follow the change of the Hubble rate. This simple
effect changes the inflationary predictions.

Let us describe the effect in some detail. Consider the inflationary dynamics driven by a scalar
field $\phi$ with potential $V(\phi)$. We now assume the presence of a heavy scalar
field $\Phi$ with mass $m$ and VEV $\Phi_0$  in the present-day vacuum. We will also assume that the 
mass of the heavy field is larger than the Hubble rate during inflation, $H\lsim m$. 
The important observation is that, in general, the inflaton field and the heavy field
are not totally decoupled from each other. On the contrary, one expects that 
during inflation the total potential assumes the form

\begin{equation}
\label{pot}
U\left(\phi,\Phi\right)=V(\phi)f\left(\Phi/M\right)+\frac{1}{2}m^2\left(\Phi-\Phi_0\right)^2,
\end{equation}
where we have expanded the potential of the heavy field around its present VEV $\Phi_0$ (assuming
consistently that $V(\phi)$ is much larger than the heavy field potential at $\Phi_0$) and the function
$f(\Phi/M)$ parametrizes the interaction between the inflaton  and the heavy 
fields. The scale $M$ may be the reduced Planck scale $M_p$ if the interaction
is of gravitational nature, but may be smaller if $f(\Phi/M)$ arises from the exchange of some
heavy fields. The kind of interaction parametrized by the function $f$ may arise, for instance, in any
model of inflation incorporated within supergravity. Indeed, radiative corrections
to the K\"{a}hler potential lead to terms in the effective Lagrangian of the form

\begin{equation}
\delta{\cal L}=\pm \frac{C^2}{M_p^2}\int d^4\theta\,\phi^\dagger\phi\,\Phi^\dagger\Phi=
\mp C^2 \frac{V(\phi)}{M_p^2}\Phi^\dagger\Phi,
\end{equation}
where $C^2$ is a coefficient which can be larger than unity.
We can now Taylor-expand the function $f(\Phi/M)$ around $\Phi_0$ to find
how much the heavy field VEV is displaced from its present-day value. The full 
potential becomes

\begin{equation}
\label{pottaylor}
U\left(\phi,\Phi\right)\simeq V(\phi)f_0+  V(\phi)\frac{f'_0}{M}(\Phi-\Phi_0)
+V(\phi)\frac{1}{2}\frac{f''_0}{M^2}(\Phi-\Phi_0)^2
+\frac{1}{2}m^2\left(\Phi-\Phi_0\right)^2,
\end{equation}
where $f_0=f(\Phi_0/M)$, $f'_0$ denotes the derivative of $f$ with respect to its argument 
evaluated at $\Phi=\Phi_0$  and so on. 
One can easily show that the VEV of the heavy field is 
shifted from $\Phi_0$ to

\begin{equation}
\Phi\simeq \Phi_0-\frac{f'_0 V(\phi)/M}{f''_0 V(\phi)/M^2+m^2}.
\end{equation}
Plugging this new VEV back into Eq. (\ref{pottaylor}) one finds the potential

\begin{equation}
\label{potential}
U(\phi)\simeq V(\phi)f_0-\frac{1}{2}\frac{\left(f'_0\right)^2 }{M^2 m^2}V^2(\phi),
\end{equation}
where we have  assumed $|f''_0|V(\phi)/M^2\lsim m^2$ for the sake of simplicity.  We obtained what  advertised earlier: 
because of the small change during inflation of the VEV of the heavy field with respect to its value
in the present vacuum, the starting inflaton  potential receives a generic correction 

 \begin{equation}
 \delta V(\phi)\sim
(V^2(\phi)/M^2 m^2)\sim (H(\phi)/m)^2(M_p/M)^2 V(\phi).
\end{equation}
Let us now compute the corresponding changes in the inflationary observables. First of all, the number
of e-folds to go till the end of inflation
becomes

\begin{equation}
\label{N}
N=\int dt'\,H(t')\simeq \frac{1}{M_p^2}\int^{\phi_N} d\phi\, \frac{V}{V'} \left(1+\frac{1}{2}
\frac{\left(f'_0\right)^2}{f_0 M^2 m^2} V\right),
\end{equation} 
where $V'=dV(\phi)/d\phi$ and so on. The slow-roll parameters become

\begin{eqnarray}
\label{sr}
\epsilon&\simeq& \frac{M_p^2}{2}\left(\frac{V'}{V}\right)^2\left(1-\frac{\left(f'_0\right)^2 V}{f_0 M^2 m^2}\right),\nonumber\\
\eta&\simeq& M_p^2\left(\frac{V''}{V}\right)\left(1-\frac{1}{2}\frac{\left(f'_0\right)^2 V}{f_0 M^2 m^2}
\right) -M_p^2\left(\frac{V'}{V}\right)^2\frac{\left(f'_0\right)^2 V}{f_0 M^2 m^2}\, ,
\end{eqnarray}
and are to be computed at $\phi=\phi_N$ when there are $N$ e-folds to go till the end of inflation. 
From these expressions we conclude
that the spectral index $n_\zeta=1+2\eta-6\epsilon$, the tensor-to scalar ratio $r$ and their running with the
scales (or e-folds) $d n_\zeta/dN$ and $dr/dN$  get corrections of the form

\begin{eqnarray}
\delta n_\zeta&=&{\cal O}\left(\epsilon,\eta\right)\frac{\left(f'_0\right)^2}{f_0}\frac{M_p^2}{M^2}\frac{H^2}{m^2},\nonumber\\
\delta r&=&{\cal O}\left(\epsilon\right)\frac{\left(f'_0\right)^2}{f_0}\frac{M_p^2}{M^2}\frac{H^2}{m^2},\nonumber\\
\delta\left(\frac{d n_\zeta}{d N}\right)&=&{\cal O}\left(\epsilon^2,\eta^2,\epsilon\eta\right)
\frac{\left(f'_0\right)^2}{f_0}\frac{M_p^2}{M^2}\frac{H^2}{m^2},\nonumber\\
\delta\left(\frac{d r }{d N}\right)&=&{\cal O}\left(\epsilon^2,\eta^2,\epsilon\eta\right)
\frac{\left(f'_0\right)^2}{f_0}\frac{M_p^2}{M^2}\frac{H^2}{m^2}.
\label{aa}
\end{eqnarray}
These corrections may be large as $H$ is not necessarily much smaller than
the mass $m$ of the heavy field, $M$ can be smaller than the Planck scale and there may be many heavy
fields.
The corrections  (\ref{aa}) have to be compared to 
the corrections which are second-order in the slow-roll parameters, {\it e.g.} 
 $\delta n_\zeta={\cal O}(\epsilon^2,\eta^2,\epsilon\eta)$. The corrections due to the
presence of the heavy field can be clearly larger. 

Consider for example the large-field models of inflation characterized by a potential
$V(\phi)=(\mu^{4-p}/p)\phi^p$ where $\mu$ is a mass scale and $p$ is a positive integer. 
 One can easily check that the spectral
 index $n_\zeta$ is not modified as a function 
of the number of e-folds $N$ given in Eq. (\ref{N}) and we find $n_\zeta-1=-(2+p)/(2N)$~\cite{comment}. 
However, the tensor-to-scalar perturbation ratio $r$ is modified. Since in single-field inflation $r=16\epsilon$, 
we can easily calculate 
\begin{equation}
r= \frac{8p}{p+2}(1-n_\zeta)\left(1- \frac{3(p+1)}{p+2}
\frac{\left(f'_0\right)^2}{f_0}\frac{M_p^2}{M^2}\frac{H^2(\phi_N)}{m^2}\right).
\end{equation}  
Notice that the correction is always negative (this in fact holds for any inflaton potential). Thus,  for example, 
 for a quartic potential, $V(\phi)\propto \phi^4$, which has been put under strong pressure from the recent data 
analyses, 
this means that a slight red tilt could still be accommodated with an amplitude of the gravity waves that is lower 
than the usual standard predictions. Also, we conclude that the corrections from the presence of the
heavy fields may be even larger than the expected accuracy $r\sim 10^{-4}$ of forthcoming experiments
aimed to measure the presence of tensor modes through the $B$-type polarization of the CMB \cite{uros}.

The running of $r$ will also change (while that of $n_\zeta$ is not affected by the
presence of the heavy field)
into

\begin{equation}
\frac{dr}{dN}= -\frac{16 p}{(p+2)^2}(1-n_\zeta)^2-\frac{3}{8}\frac{(p-2)(p+1)}{p(p+2)}\,r^2
\frac{\left(f'_0\right)^2}{f_0}\frac{M_p^2}{M^2}\frac{H^2(\phi_N)}{m^2}.
\end{equation} 
Notice that for $p\gg 1$, the running of $r$ is due only to the heavy field correction.
As another  example, let us consider an inflaton potential of the form
\begin{equation}
\label{log}
V(\phi)=V_0\left(1+\alpha\ln \frac{\phi}{\Lambda}\right),
\end{equation}
which is typical  in supersymmetric hybrid inflation~\cite{lrreview} (here $\alpha$ is usually a loop-factor and 
$\Lambda$ is a mass scale). 
After some simple algebra, 
one finds that

\begin{equation}
n_\zeta-1=-\frac{1}{N}\left(1+\frac{3}{2} \alpha\right)-\frac{3}{4}
\frac{\alpha}{N}\frac{\left(f'_0\right)^2}{f_0 }\left(\frac{M_p}{M}\right)^2\left(\frac{H}{m}\right)^2.
\end{equation}
The almost unavoidable presence of heavy fields during inflation
leads therefore to corrections to the inflationary observables which may be larger than both  the second-order 
contributions in the slow-roll parameters and of the accuracy of future 
experiments. 
Indeed, while $H\lsim m$ for the field $\Phi$ to be considered
heavy during inflation, one might have easily $M\lsim M_p$ and a large contribution from the $(f'_0)^2/f_0$. 
Moreover, while here we have just considered the case of a single heavy scalar field, the size of these 
effects could be increased by the presence of many  of such heavy fields (for instance, in inflationary 
string-motivated scenarios one has plenty of moduli fields, such as the complex structure moduli, which
are usually disregarded). In a well-defined model of inflation, well-rooted
in a given particle theory model, these corrections can (and should) be computed. On the other hand,
if one wishes to see if a generic (toy) inflaton potential is in agreement with present and future
observations, these corrections should be regarded as an unavoidable source of uncertainty
in the theoretical predictions. This is rather unfortunate. However, one could also
try to make use of these corrections to gain something. For instance, since the
corrections to $r$, $\epsilon$ and $\eta$ are generically negative, one could try to relax the
so-called Lyth bound on the total variation of the inflaton field \cite{bound} and to obtain 
inflation from potentials which are too steep.

\acknowledgments
\noindent
This research was supported in part by the European Community's Research
Training Networks under contracts MRTN-CT-2004-503369 and MRTN-CT-2006-035505. N.B. acknowledges the kind hospitality 
of the Theory Divison of CERN where part of this work has been carried out.


\begin{thebibliography}{99}

\bibitem{lrreview} For a review, see
D.~H.~Lyth and A.~Riotto,
%``Particle physics models of inflation and the cosmological density
%perturbation,''
Phys.\ Rept.\  {\bf 314}, 1 (1999). 
%%CITATION = HEP-PH 9807278;%%


\bibitem{wmap}
  D.~N.~Spergel {\it et al.}  [WMAP Collaboration],
  %``Wilkinson Microwave Anisotropy Probe (WMAP) three year results:
  %Implications for cosmology,''
  Astrophys.\ J.\ Suppl.\  {\bf 170}, 377 (2007)
  [arXiv:astro-ph/0603449].
  %%CITATION = APJSA,170,377;%%




\bibitem{planck} See {\tt http://planck.esa.int/}.






\bibitem{KK1}
  W.~H.~Kinney, E.~W.~Kolb, A.~Melchiorri and A.~Riotto,
  %``Inflation model constraints from the Wilkinson microwave anisotropy  probe
  %three-year data,''
  Phys.\ Rev.\  D {\bf 74}, 023502 (2006)
  [arXiv:astro-ph/0605338].
  %%CITATION = PHRVA,D74,023502;%%

\bibitem{martin} J.~Martin and C.~Ringeval,
  %``Inflation after WMAP3: Confronting the slow-roll and exact power  spectra
  %to CMB data,''
  JCAP {\bf 0608}, 009 (2006)
  [arXiv:astro-ph/0605367].

\bibitem{finelli} F.~Finelli, M.~Rianna and N.~Mandolesi,
  %``Constraints on the Inflationary Expansion from Three Year WMAP, small scale
  %CMB anisotropies and Large Scale Structure Data Sets,''
  JCAP {\bf 0612}, 006 (2006)
  [arXiv:astro-ph/0608277].







\bibitem{Lidseyetal}
See, for example, 
  J.~E.~Lidsey, A.~R.~Liddle, E.~W.~Kolb, E.~J.~Copeland, T.~Barreiro and M.~Abney,
  %``Reconstructing the inflaton potential: An overview,''
  Rev.\ Mod.\ Phys.\  {\bf 69}, 373 (1997)
  [arXiv:astro-ph/9508078].
  %%CITATION = RMPHA,69,373;%%



\bibitem{comment} It is worth mentioning that there might be specific cases when the corrections to the spectral index 
cancel each other thus leaving the 
expression of the spectral index unchanged as a function of the number of e-folds $N$. 
 Taking into account the corrections appearing in Eqs.~(\ref{N})-(\ref{sr}), the slow-roll parameters can be expressed a
$\epsilon=\epsilon(N)(1+\delta\epsilon(N))$ (and the same for $\eta$), where $\epsilon(N)$ and $\eta(N)$ are parametrically 
the same functions of the 
number of e-folds as the ones predicted withouth any correction to the potential $V(\phi)$. In other words, 
$\epsilon(N)=\epsilon_0(N_0 \rightarrow N)$, where the subscript 0 stands for the quantities computed withouth invoking 
any corrections. It is thus easy to 
see that the form of the spectral index $n_\zeta=1+2\eta-6\epsilon$  is left unchanged as a function of $N$ if the following 
condition is satisfied $
\delta \eta(N)/\delta \epsilon(N)=3\epsilon_0(N_0)/\eta_0(N_0)$.





\bibitem{uros} M.~Amarie, C.~Hirata and U.~Seljak,
  %``Detectability of tensor modes in the presence of foregrounds,''
  Phys.\ Rev.\  D {\bf 72}, 123006 (2005) [arXiv:astro-ph/0508293];
U.~Seljak and C.~M.~Hirata,
   %``Gravitational lensing as a contaminant of the gravity wave signal in
   %CMB,''
   Phys.\ Rev.\  D {\bf 69}, 043005 (2004)
   [arXiv:astro-ph/0310163]
   %%CITATION = PHRVA,D69,043005;%%
;  F.~Stivoli, C.~Baccigalupi, D.~Maino and R.~Stompor,
   %``Separating cosmological B modes from foregrounds in cosmic microwave
   %background polarization observations,''
   Mon.~Not.~R.~astr.~Soc.~{\bf 372}, 615 (2006),
   [arXiv:astro-ph/0505381].
   %%CITATION = ASTRO-PH/0505381;%%



\bibitem{bound}  D.~H.~Lyth,
  %``What would we learn by detecting a gravitational wave signal in the  cosmic
  %microwave background anisotropy?,''
  Phys.\ Rev.\ Lett.\  {\bf 78}, 1861 (1997)
  [arXiv:hep-ph/9606387].
\end{thebibliography}
\end{document}